\documentclass[twocolumn]{aastex62}
\usepackage{natbib}
\usepackage{hyperref}

\usepackage{graphicx,amsmath,amssymb,xspace,here}
\usepackage{color}

\newcommand{\targ}{C21434\xspace}

\newcommand{\Mgas}{\ensuremath{M_{\rm{H_2}}}\xspace}
\newcommand{\Mstar}{\ensuremath{M_{\rm{\star}}}\xspace}

\newcommand{\fgas}{\ensuremath{f_{\rm{H_2}}}\xspace}

\newcommand{\msol}{\ensuremath{\rm{M}_\odot}\xspace}

\newcommand{\lprime}{\ensuremath{\rm{L}_{\rm{CO}}'}\xspace}
\newcommand{\arc}{\ensuremath{''}\xspace}
\newcommand{\um}{\ensuremath{\mu\rm{m}}\xspace}
\newcommand{\kms}{\ensuremath{\rm{km\,s}^{-1}}\xspace}
\newcommand{\uJy}{\ensuremath{\mu\rm{Jy}}\xspace}

\newcommand{\alphaco}{\ensuremath{\alpha_{\rm{CO}}}\xspace}

\begin{document}

\title{Extremely Low Molecular Gas Content in a Compact, Quiescent Galaxy at $z=1.522$}

\author[0000-0001-5063-8254]{Rachel Bezanson}
\affiliation{Department of Physics and Astronomy and PITT PACC, University of Pittsburgh, Pittsburgh, PA, 15260, USA}

\author[0000-0003-3256-5615]{Justin Spilker}
\affiliation{Department of Astronomy, University of Texas at Austin, 2515 Speedway, Stop C1400, Austin, TX 78712, USA}

\author[0000-0003-2919-7495]{Christina C. Williams}
\affiliation{Steward Observatory, University of Arizona, 933 North Cherry Avenue, Tucson, AZ 85721, USA}
\affiliation{NSF Fellow}

\author[0000-0001-7160-3632]{Katherine E. Whitaker}
\affiliation{Department of Physics, University of Connecticut, 2152 Hillside Road, Unit 3046, Storrs, CT 06269, USA}

\author[0000-0002-7064-4309]{Desika Narayanan}
\affiliation{Department of Astronomy, University of Florida, 211 Bryant Space Science Center, Gainesville, FL 32611, USA}

\author[0000-0001-6065-7483]{Benjamin Weiner}
\affiliation{Steward Observatory, University of Arizona, 933 North Cherry Avenue, Tucson, AZ 85721, USA}

\author[0000-0002-8871-3026]{Marijn Franx}
\affiliation{Leiden Observatory, Leiden University, P.O.Box 9513, NL-2300 AA Leiden, The Netherlands}


\begin{abstract}

One of the greatest challenges to theoretical models of massive galaxy formation is the regulation of star formation at early times. The relative roles of molecular gas expulsion, depletion, and stabilization are uncertain as direct observational constraints of the gas reservoirs in quenched or quenching galaxies at high redshift are scant.  We present ALMA observations of CO(2--1) in a massive ($\log \Mstar/\msol=11.2$), recently quenched galaxy at $z=1.522$. The optical spectrum of this object shows strong Balmer absorption lines, which implies that star formation ceased $\sim$0.8\,Gyr ago. We do not detect CO(2--1) line emission, placing an upper limit on the molecular $\mathrm{H_2}$ gas mass of 1.1$\times10^{10}\,$\msol. The implied gas fraction is $\fgas{\equiv M_{H_2}/M_{\star}}<7\%$, $\sim10\times$ lower than typical star forming galaxies at similar stellar masses at this redshift, among the lowest gas fractions at this specific star formation rate at any epoch, and the most stringent constraint on the gas contents of a $z>1$ passive galaxy to date. Our observations show that the depletion of $\mathrm{H_2}$ from the interstellar medium of quenched objects can be both efficient and fairly complete, in contrast to recent claims of significant cold gas in recently quenched galaxies.  We explore the variation in observed gas fractions in high-$z$ galaxies and show that galaxies with high stellar surface density have low \fgas, similar to recent correlations between specific star formation rate and stellar surface density. 

\end{abstract}

\section{Introduction}

Producing realistic populations of non-star forming or quiescent
 galaxies over cosmic time remains a significant challenge to current theoretical models of galaxy formation and evolution. Quenched galaxies have been identified as early as $z\sim4$ \citep[e.g.][]{straatman:14,glazebrook:17},
but their emergence peaks at $z\sim2$, an epoch after which the majority above $\log{\rm \Mstar}/\msol\gtrsim11$ have their star formation truncated \citep[e.g.,][]{muzzin:13,tomczak:14,davidzon:17}. The physical mechanisms responsible for rapidly halting the star formation in early massive galaxies, and preventing future star formation for many Gyr, remain insufficiently constrained, partly  
due to a poor understanding of the observable signatures of the physics affecting star formation. These processes are likely tied to either depleting, expelling, and/or heating the cold molecular gas  
in galaxies, which would otherwise fuel star formation \citep[see e.g.,][and references therein]{man:18}.

This has motivated investigations into the molecular gas properties of quenched galaxies, as probed by the rotational transitions of CO. The overwhelming majority of these studies have, until recently, been limited to the local Universe, where observations indicate very low molecular gas fractions ($0.1-1\%$) and very low star formation efficiency relative to star forming galaxies \citep[e.g.,][]{saintonge:11,saintonge:11b, saintonge:12, davis:11, davis:13}. Locally, massive quiescent galaxies have old stellar populations, indicating that their primary epoch of star formation occurred many Gyrs in the past; the residual molecular gas reservoirs are likely either supplied by external processes such as gas-rich merging \citep[e.g.,][]{young:14} or internally via stellar mass loss \citep[e.g.,][]{davis:16}.  Late accretion of gas will often be characterized by misaligned stellar and molecular gas kinematics, which may not be as important even as late as $z\sim0.7$ \citep{hunt:18}. However, recently quenched (e.g., post-starburst) galaxies, which have ceased star formation within the last $\lesssim1$Gyr, provide the opportunity to observe the molecular gas properties 
immediately following a recent star forming episode to gain better insight into the quenching process. Observations of cold gas in local post-starburst galaxies indicate large ($\sim 10^9 M_{\odot}$ or $\fgas\sim 0.01-0.3$) reservoirs of cold gas despite their low star formation rates \citep[e.g.][]{french:15, french:18}, suggesting that the recent quenching of these galaxies cannot be simply due to a lack of cold gas.  These transitioning galaxies are extraordinarily rare in the local Universe, $<$0.2\% of the overall population and rarer at
high masses \citep[e.g.,][]{french:15}, but become far more common at $z>1$ \citep{whitaker:12a}. 

Exploring the molecular gas content of high-redshift transitioning galaxies is necessary to establish which physical mechanisms are responsible for building up the massive end of the red sequence. Outside of the local Universe only ${\sim}10$ $z<1$ quiescent galaxies have published constraints on their molecular gas reservoirs based on CO lines \citep{suess:17,spilker:18}, indicating a large spread of gas fractions ($\fgas\equiv\Mgas/M_{\star}$) from upper limits of $\sim3\%$ to measured $f_{H_2}\sim15\%$ in galaxies below the star forming main sequence \citep{spilker:18} and between 4--20\% for massive post-starburst galaxies \citep{suess:17}. 
Measurements beyond $z\sim1$ are even more sparse, with one quiescent galaxy at $z\sim1.4$ constrained to $f_{H_2}<10\%$ \citep{sargent:15}, two cluster galaxies at $z=1.46$ \citep{hayashi:18} and $z\sim1.62$ \citep{rudnick:17} with $f_{H_2}=35\%$ and $\sim42\%$ respectively.\footnote{Using the stellar mass estimate from \citet{skelton3dhst} 3D-HST catalogs for maximal consistency, which is $\sim$0.3 dex lower than the value quoted in \citet{rudnick:17}, corresponding to $f_{H_2}=20\%$. This discrepancy is slightly larger than the $\sim$0.2 dex uncertainty expected for stellar mass estimates with a fixed IMF \citep{muzzin:09}.}   
In contrast with the \citet{sargent:15} measurement, but perhaps consistent with the spread in the small number of observed individual galaxies, \citet{gobat:18} performed a stacking analysis of dust continuum and determined that quiescent galaxies at $z\sim1.8$ have average gas fractions $\sim16\%$. 
Although these results are not in tension, there is much work to be done to define the distribution and scatter in cold gas reservoirs remaining in galaxies as they shutdown star formation at this key epoch.
With this work, we quantify the molecular gas reservoir in \targ, a massive and recently quenched galaxy at $z=1.52$, leveraging the unrivaled sensitivity of ALMA at the peak of galaxy assembly. Our ALMA observations represent the deepest constraint to date on the molecular gas content of a non-starforming galaxy at $z>1$. We assume a standard concordance cosmology with $\mathrm{H_0=70 km\,s^{-1}Mpc^{-1}}$, $\Omega_M=0.3$, $\Omega_{\Lambda}=0.7$, and a \citet{chabrier:03} Initial Mass Function.

\begin{figure*}[!t]
    \centering
    \includegraphics[width=\textwidth]{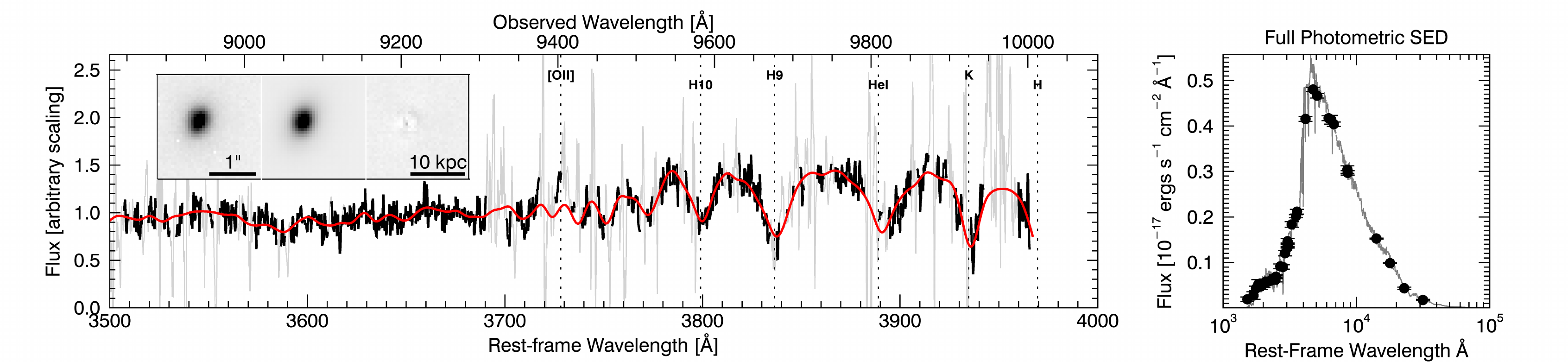}
    \caption{Left Panel: Keck/LRIS optical spectrum (black) and best-fitting model (red) and inset HST-WFC3 F160W image, S\'ersic model, and residual. Right panel: Photometric spectral energy distribution (SED) for C21434 from the NMBS photometry. The strong Balmer absorption features and peaked SED reflect the young, quiescent stellar population of this galaxy.}
    \vspace{+20pt}
    \label{fig:LRISspectrum}
\end{figure*}

\section{Target Selection and Data}

\subsection{Multi-Wavelength Photometry and LRIS Spectrum}

The targeted galaxy, C21434, is selected from a sample of massive galaxies ($\log\,\Mstar/\msol\,\gtrsim11$) at $z\sim 1.5$ in the NEWFIRM Medium Band Survey \citep[NMBS,][]{whitaker:11}. The sample and full analysis is described in \citet{bezanson:13a}, but we briefly summarize here. The NMBS survey provides extensive multi-wavelength photometry for this object from the UV to 24\,\um, including medium-band near-IR filters that span the Balmer/$4000\mathrm{\AA}$ break at the redshift of the target (shown in the right panel of Figure \ref{fig:LRISspectrum}). A deep (18 hour) optical spectrum was taken in January and April 2010 using the LRIS spectrograph (main panel of Figure \ref{fig:LRISspectrum}) and high-resolution rest-frame optical imaging was obtained using the $F160W$ filter on the HST-WFC3 Camera (Program HST-GO-12167, PI: Franx). The structural parameters of C21434 are measured using \textit{Galfit} \citep{galfit} to fit a single \citet{sersic} profile to the HST image. The image, best-fitting model, and residuals from the fit are included in the inset of Figure \ref{fig:LRISspectrum}.

The NMBS photometry are fit using FAST \citep{kriek:09} with \citet{bc:03} stellar population synthesis models 
and delayed exponentially declining star formation histories, fixing to the spectroscopic redshift $z_{spec}=1.522$.  The best-fitting model finds a stellar mass of $\log \Mstar/\msol = 11.2$. The strongly peaked Balmer break in the SED and strong Balmer absorption features indicate that C21434 is a recently quenched ``A-type'' post-starburst galaxy. The photometric fit yields a stellar age of $\sim$0.8\,Gyr. We also spectroscopically fit the age of C21434 using PPXF \citep{cappellari:04} to fit a linear, non-negative sum of \citet{vazdekis:99} single stellar population models, again finding a stellar age of $\sim$0.8\,Gyr.

\subsection{ALMA Observations} \label{obs}

ALMA Band 3 observations were carried out in two consecutive observing blocks on 2016 January 18 as part of project 2015.1.00853.S (PI: Bezanson). The CO(2--1) line is redshifted to a sky frequency of 91.411\,GHz.  One 1.875\,GHz spectral window was centered around this frequency with 7.8\,MHz ($\sim$25\,\kms) channelization, with two additional 1.875\,GHz spectral windows placed for continuum observations. The total duration of the two blocks was 160\,min, with 109\,min on source.  A total of 40 antennas were active in the array, reaching maximum baselines of 330\,m, yielding a resolution of $\sim$2\arc. Quasars J1058+0133 and J0949+0022 served as the bandpass and complex gain calibrators, respectively, for both observing blocks. J1058+0133 was also used to calibrate the absolute flux scale in the first observing block, while Ganymede served this purpose in the second block. We verified that the flux density of the gain calibrator was consistent between the two tracks to within 3\%.  The data were reduced using the standard ALMA Cycle 3 pipeline, and no significant issues with this reduction were found. The reduced data reach a continuum sensitivity of 9\,\uJy/beam at 98\,GHz, and a CO(2--1) line sensitivity of 105\,\uJy/beam per 100\,\kms channel. 
\begin{figure}
    \includegraphics[width=0.95\columnwidth]{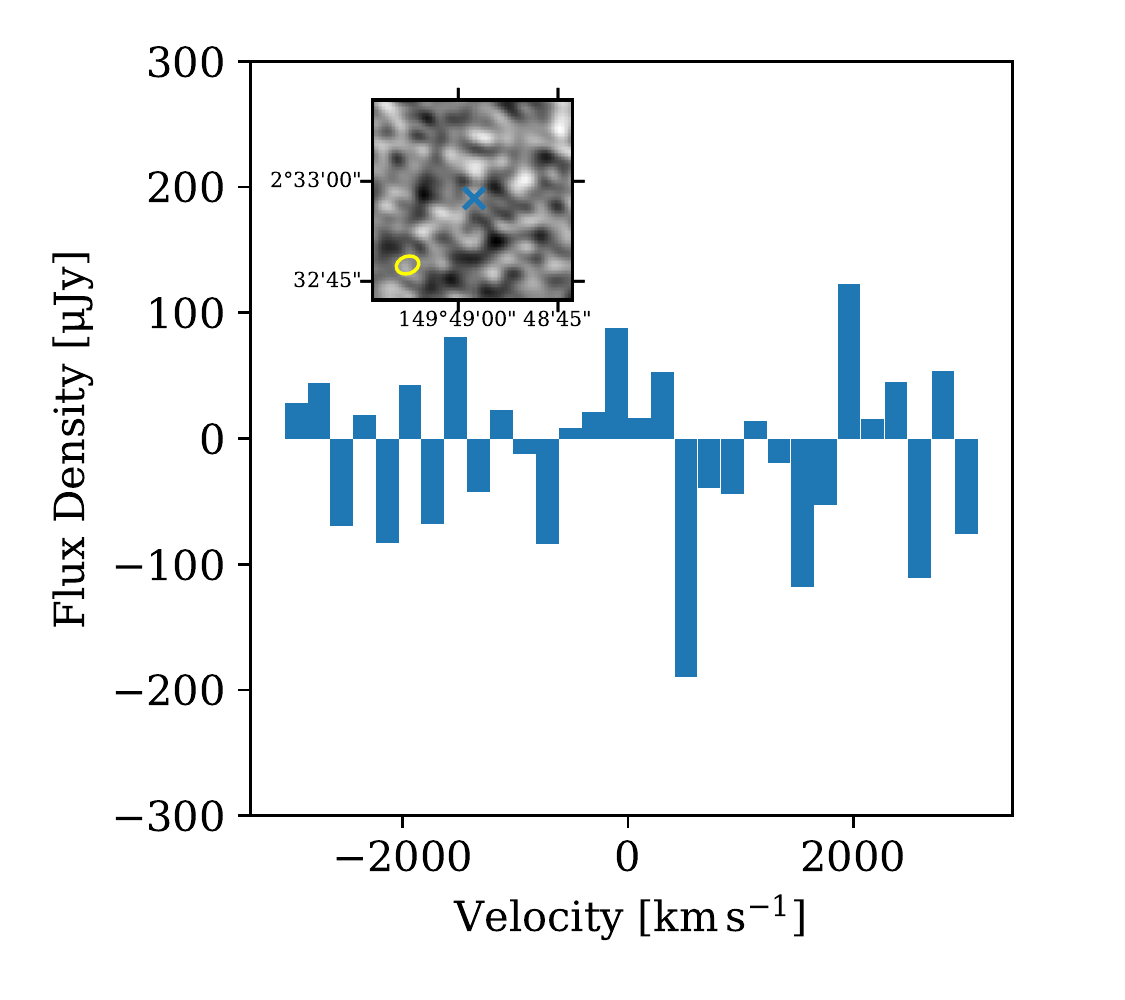}
    \caption{CO(2--1) spectrum {and inset CO(2-1) map} demonstrating the non-detection of \targ. {The CO(2-1) map is integrated within 600 $\mathrm{km\,s^{-1}}$ of the optical redshift and the optical position of \targ is indicated by a blue cross.} We place a 3$\sigma$ upper limit on the molecular gas mass of $\Mgas<1.1\times10^{10}$\,\msol, corresponding to a molecular gas fraction of $<7\%$.}
    \label{fig:spectrum}
\end{figure}

\subsection{Non-Detection and $H_2$ Gas Mass Limit} \label{mgas}

To extract a spectrum of \targ, we fit a point source to the visibilities in bins of 8 channels, or $\sim$200\,\kms. The stellar effective radius, 0.23\arc, is much smaller than the resolution of these observations, $\sim$2\arc, and so this unresolved source is expected to be pointlike. We fix the position of the modeled point source to its position in optical/NIR imaging, leaving only the flux density at each velocity channel as a free parameter. The extracted spectrum is shown in Fig.~\ref{fig:spectrum}. We do not detect \targ in either CO(2--1) emission or 3\,mm continuum.
\begin{figure*}
    \centering
    \includegraphics[width=0.99\textwidth]{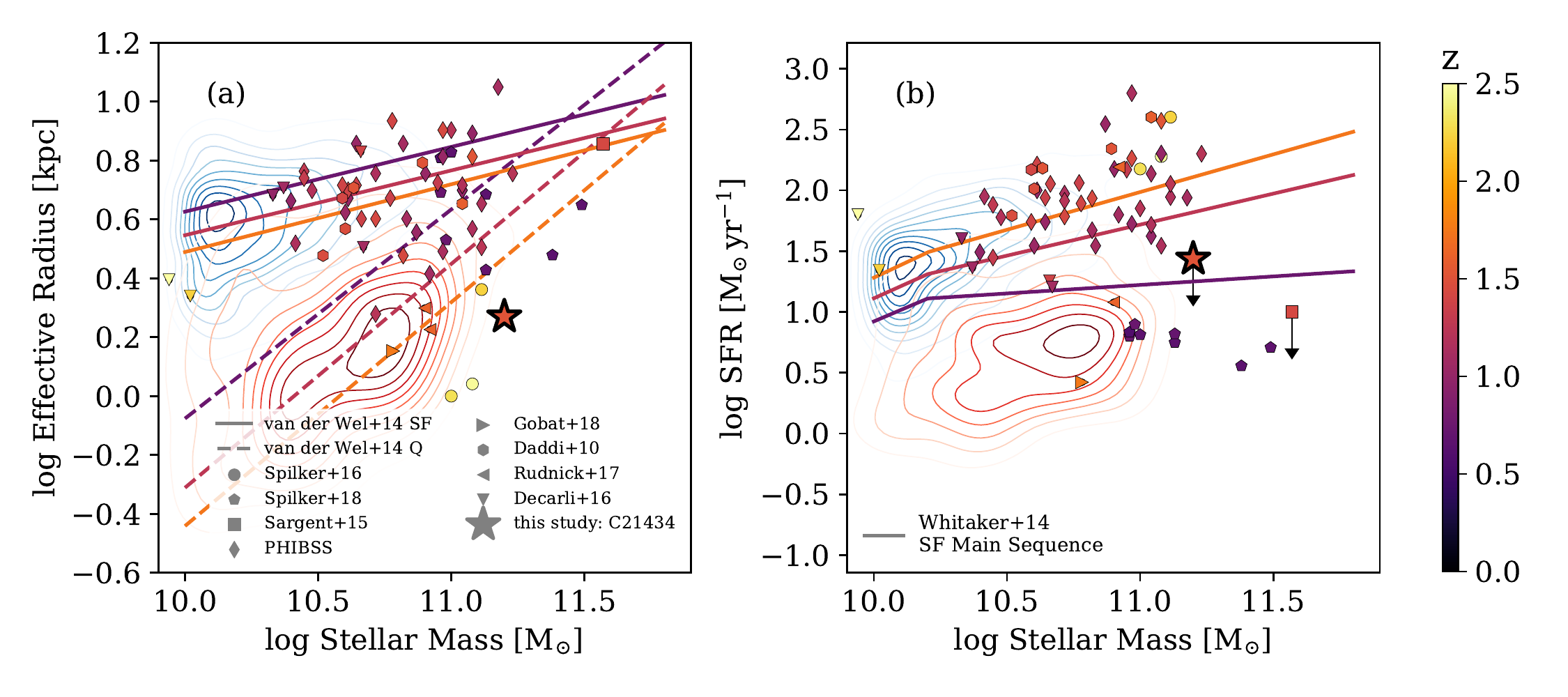}
    \caption{Galaxy size (left panel) and star formation rates (right panel) as a function of stellar mass. Contours indicate the location of star forming (blue contours) and quiescent (red contours) galaxies at $1<z<2$ in the 3D-HST survey. \citet{wel:14} size-mass relations at several epochs (left) and \citet{whitaker:14} star formation ``main sequence'' (right) are included as a solid and dashed lines for star forming or quiescent relations respectively. Galaxies from a variety of samples with molecular gas and rest-frame optical size measurements are indicated by colored symbols \citep{daddi:10,tacconi:10,sargent:15,decarli:16,spilker:16,rudnick:17,gobat:18,spilker:18}, with \targ indicated by a red star. Symbol and line colors indicate the redshift of each object or relation. The majority of targeted galaxies have been limited to the extended (left) and highly star forming population (right), however, a few studies have begun to probe the reservoirs of galaxies like \targ with compact structures \citep{spilker:16} and and/or minimal on-going star formation \citep{sargent:15,rudnick:17,gobat:18}. }
    \label{fig:selections}
\end{figure*}


To set an upper limit on the integrated CO(2--1) flux of \targ, we assume that the gas spatial and kinematic distribution traces the stellar continuum and therefore adopt a CO line width equal to the line width determined from the stellar absorption features, which have FWHM $\sim$ 600\,\kms \citep{bezanson:13a}. This yields a 3$\sigma$ upper limit to the CO(2--1) luminosity of $\lprime < 2.4 \times 10^{9}$\,K\,\kms\,pc$^2$.

In order to convert this limit to a molecular gas mass, we assume thermalized line emission, as observed in local early type galaxies from the ATLAS$^{\mathrm{3D}}$ survey \citep{young:11}. Because the CO(2--1) line is near the ground state, only a limited range of CO(2--1)/CO(1--0) excitation variations are observed in galaxies; our estimate of thermalized excitation should be accurate to $<$30\%. We adopt a Milky Way-like CO-H$_2$ conversion factor, $\alphaco=4.4$\,\msol/(K\,\kms\,pc$^2$). Aside from being a conservative choice, this value is motivated by observations of local quiescent galaxies \citep{young:11} and theoretical models of the variations of \alphaco with metallicity \citep{feldmann:12,narayanan:12}. Adopting this \alphaco yields a final 3$\sigma$ upper limit on the molecular gas mass of \targ of $\Mgas < 1.1\times10^{10}$\,\msol, however this assumption will always introduce a source of systematic uncertainty. 

{We note that the non-detection of \targ from the 3\,mm continuum is far less constraining. Based on dust emissivity from \citet{dunne:11} with a dust temperature of 25K and $M_{H_2}/M_{dust}$ = 100, the 3-$\sigma$ gas mass limit from the 3\,mm continuum is $M_{H_2}<4.5\times10^{10} M_{\odot}$, which is significantly less constraining than the CO(2-1) line flux. This places a very weak limit on $\alpha_{CO} < 19$\,\msol/(K\,\kms\,pc$^2$).}

\subsection{Ancillary Datasets} \label{ancillary}

In addition to the new ALMA observations presented herein, we compile a literature sample of high-redshift galaxies  with measured molecular gas reservoirs, star formation rates, and stellar sizes measured from rest-frame optical HST imaging. We include 38 star forming galaxies from PHIBSS \citep[CO(3--2),][]{tacconi:10,tacconi:13}, 97 (38 at $z>1$) from PHIBSS2 \citep[][]{tacconi:18} and five from \citet[][]{daddi:10} with PdBI observations of CO(2--1). We note that for PHIBSS2 we have no additional information about the uncertainty or methodology(ies) used to measure effective radii \citep{tacconi:18}. In the absence of information about e.g., how well-fit these galaxies are by S\'ersic profiles, whether they are fit in the rest-frame optical from high-resolution HST imaging, and whether sizes are circularized, we exclude the sample from structural comparisons.  
We include seven galaxies from \citet{decarli:16} with robust (flag=0) rest-frame optical HST size measurements \citep{wel:12} and reliable stellar masses ($\log M_{\star,3D-HST}/M_{\odot} > 9$), correcting to a Milky Way $\alphaco$. Three compact star forming galaxies with CO(1-0) VLA observations and HST/WFC3 imaging, are included from \citet{spilker:16}. \citet{sargent:15}, \citet{hayashi:18}, and \citet{rudnick:17} represent the only other three CO-based constraints on molecular gas in high-redshift quiescent galaxies at $z=1.4277$, $z=1.451$, and $z=1.62$. We include the  \citet{sargent:15} galaxy adopting the \citet{onodera:12} stellar mass, which also uses a \citet{chabrier:03} IMF. For maximum consistency, we include 3D-HST stellar masses \citep{skelton3dhst} for the \citet{rudnick:17} galaxies. \citet{gobat:18} reported a detection of $f_{H_2}\sim16\%$  in quiescent galaxies at $z\sim1.8$, correcting to a \citet{chabrier:03} IMF using a factor of 0.55 based on \citet{longhetti:09}, obtained via a median stacking analysis of mid-IR, far-IR, sub-mm, and radio observations to study the average dust-continuum emission from 977 quiescent galaxies. 
We adopt the effective radius from the \citet{wel:14} quiescent size-mass relation at the average mass ($\langle\Mstar\rangle=6\times10^{10}M_{\odot}$) for this sample. At intermediate redshift ($z{\sim}0.6$), \citet{suess:17} present two massive post-starburst galaxies with significant molecular gas reservoirs, however the stellar sizes are unconstrained for these galaxies. We also include 8 quiescent galaxies at $z\sim0.8$ from \citet{spilker:18}.

\begin{figure*}[t]
    \centering
    \includegraphics[width=0.99\textwidth]{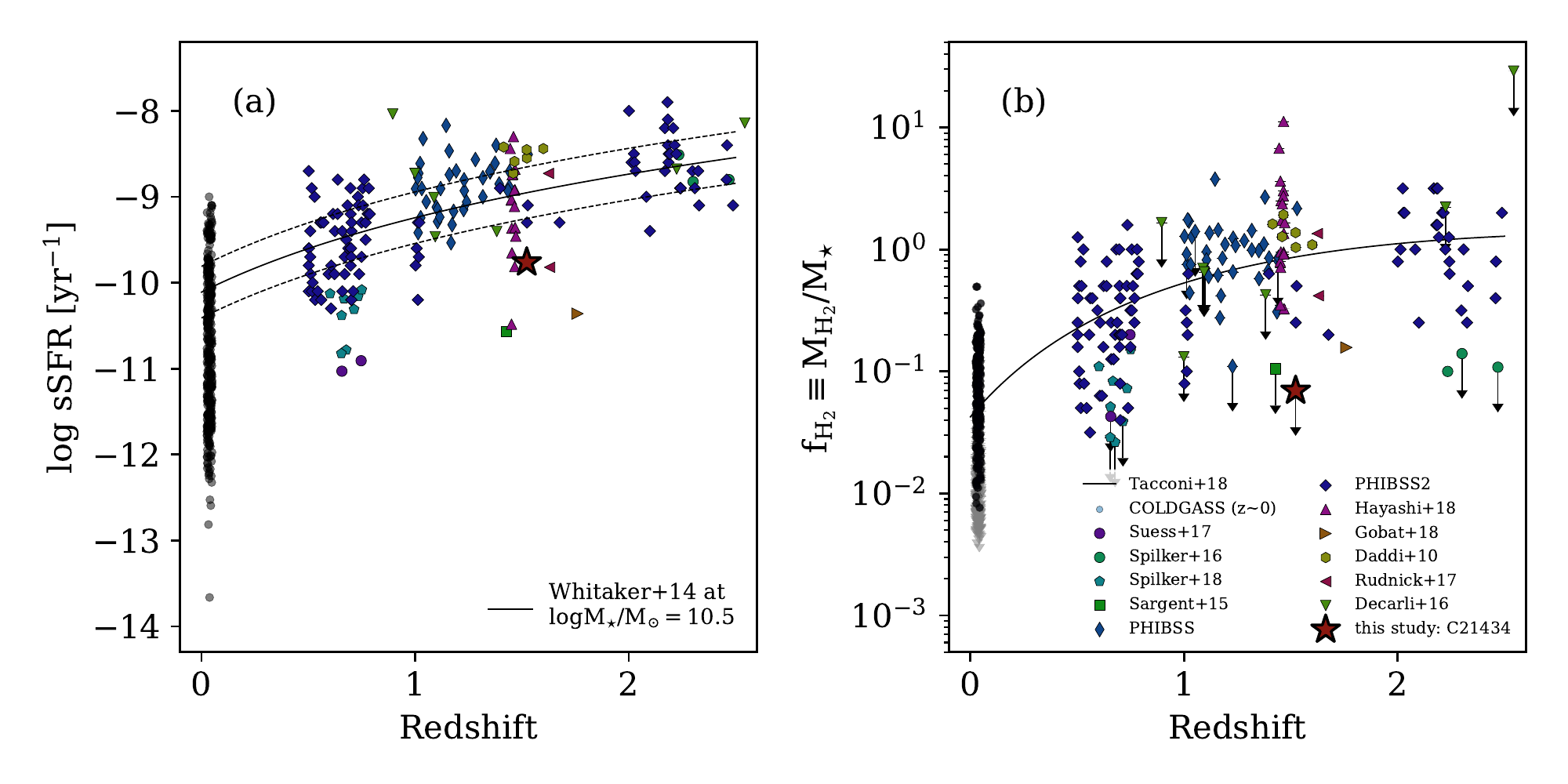}
    \caption{Specific star formation rate (sSFR, left) and molecular gas fraction ($f_{H_2}$, right) versus redshift for the high redshift sample as well as targets from the nearby COLDGASS survey \citep{saintonge:11}, which span the full range of galaxy demographics at $z\sim0$. Expected redshift evolution in sSFR \citep[left,][]{whitaker:14} and in \fgas  \citep[right,][]{tacconi:18} at $\mathrm{\log\,M_{\star}/M_{\odot}\sim10.5}$ are indicated by black lines. In the left panel, quiescent targets dramatically stand out from star forming counterparts in sSFR, however in gas fractions the \citet{spilker:16} compact star forming galaxies have similarly depleted molecular gas reservoirs.}
    \label{fig:fgas_z}
\end{figure*}

Finally we include CO(1-0)-based data from the COLDGASS survey \citep{saintonge:11, saintonge:11b,saintonge:12} as a low-redshift benchmark. For comparison to $1<z<2$ galaxies, we include galaxies from the 3D-HST photometric catalogs \citep{brammer:11,skelton3dhst}, with UV+IR star formation rates \citep{whitaker:14} and structural parameters derived from HST/WFC3 imaging \citep{wel:12}. We separate the 3D-HST galaxies into star forming and quiescent populations using  \citet{whitaker:12a} rest-frame U-V and V-J color cuts.

\section{$\mathrm{H_2}$ reservoirs across the galaxy population}

Figure \ref{fig:selections} shows \targ (red star) and the literature sample in effective radius and star formation rate as a function of stellar mass, colored by redshift. The vast majority of galaxies with measured gas reservoirs are extended (left) and star forming (right), consistent with the overall distribution of star forming galaxies at this redshift. We note that \citet{spilker:16} targets are selected to be structurally similar to compact quiescent galaxies like \targ. The \citet{sargent:15} target is sufficiently massive that it overlaps with the star forming population due to the steep quiescent size-mass relation \citep[e.g.,][]{mowla:18}.
The \citet{gobat:18} sample is clearly the least star forming $z>1$ sample, but due to its stacked nature its size is assumed on average. The full sample is heterogeneous in redshift, however we see the expected trend that galaxies at higher-$z$ (lighter symbols) are more compact (Figure \ref{fig:selections}a) and have higher SFRs (Figure \ref{fig:selections}b) at fixed mass.

Locally, there is a strong correlation between molecular gas supply and the efficiency of star formation \citep[e.g.,][]{saintonge:11}. In Figure \ref{fig:fgas_z} we show the specific star formation rate ($sSFR\equiv\,SFR/\Mstar$) and molecular gas fractions ($f_{H_2}\equiv\Mgas/\Mstar$) as a function of redshift. For reference we include the evolution of sSFR at $\Mstar\sim10^{10.5}$ \citep{whitaker:14} as a black line in Figure \ref{fig:fgas_z}a and the \citet{tacconi:18} redshift evolution (with $\log(1+z)^2$ scaling) for a star forming $\Mstar\sim10^{10.5}$ galaxy in Figure \ref{fig:fgas_z}b. In this projection, the quiescent galaxies stand out (galaxies with low sSFR in Figure \ref{fig:fgas_z}a), along with the compact star forming galaxies \citep[yellow circles,][]{spilker:16} in Figure \ref{fig:fgas_z}b, as deficient in molecular gas. We note that this comparison is relative to coeval galaxies; the molecular gas fractions of all ``depleted'' galaxies galaxies at high-redshift are still consistent with many local gas-rich star forming galaxies.

\begin{figure*}
    \centering
    \includegraphics[width=0.99\textwidth]{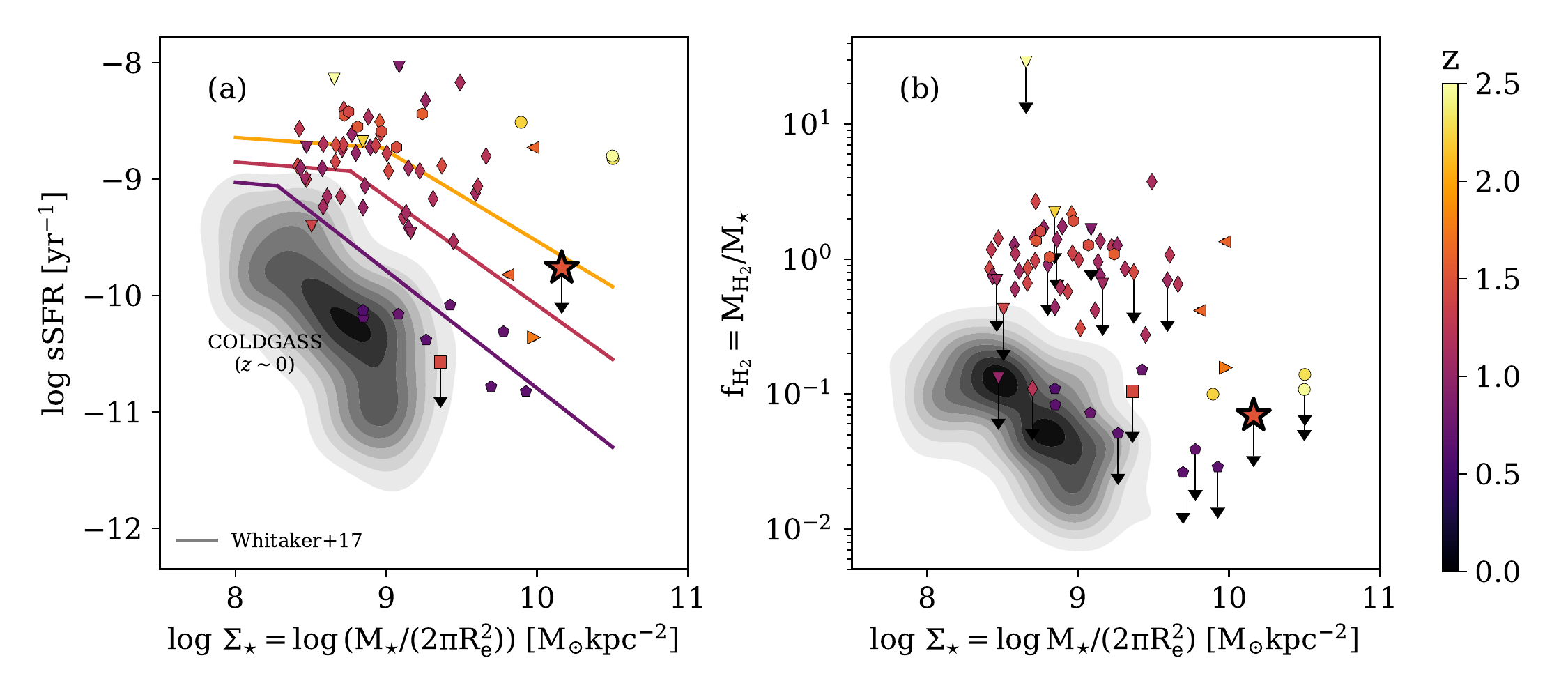}
    \caption{Specific star formation rate and $H_2$ gas fraction versus stellar mass surface density for galaxies at $z\sim0$ (gray contours, COLDGASS survey) and at higher redshift (symbols, colored by redshift). Local galaxies lie offset in sSFR (left) and \fgas (right) at fixed stellar density, as expected. Broken power law relations between sSFR and $\mathrm{\Sigma_{\star}}$ for all galaxies at $\langle z\rangle=0.75,1.25,$ and $1.75$ from \citet{whitaker:17} are indicated by solid lines in panel (a). 
    Although some galaxies at fixed density and redshift overlap in sSFR (a), gas fractions at all redshifts drop dramatically for the most dense galaxies ($\Sigma_* \gtrsim 5\times10^{9} {\rm M_{\odot} kpc^{-2}}$, b).}
    \label{fig:sigmastar}
\end{figure*}

The similarity between the molecular gas reservoirs in compact star forming and quiescent galaxies suggests a connection between stellar structures and gas fractions. This may be expected from the observed correlations between star formation and galaxy structure observed at all redshifts \citep[e.g.][and references therein]{franx:08,whitaker:17}. Figure \ref{fig:sigmastar}a shows sSFR and $f_{H_2}$ versus stellar mass surface density ($\Sigma_{\star}\equiv\,\Mstar/(2\pi R_e^2)$) where galaxies with lower sSFR tend to have higher densities at any epoch. The average relations for all galaxies in the 3D-HST survey at several epochs are included from \citet{whitaker:17}. Remarkably, although the high $\Sigma_{\star}$ populations are still mildly overlapping in sSFR with extended galaxies, they separate relatively cleanly in gas fraction (Figure \ref{fig:sigmastar}b). This trend exists at each epoch such that less dense galaxies also have higher \fgas. {We note although this appears to be at tension with recent results from \citet{freundlich:19}, which found no trends in gas mass with stellar surface density, that study is primarily based on extended star forming galaxies with much lower stellar densities; above $\Sigma_{\star}\gtrsim9.5$ those data also show hints of depleted $M_{H_2}$}.


\section{Discussion}

Measurements of the molecular gas contents of quenching galaxies provide critical constraints on the physics driving future star formation. Our deep ALMA observation provides the most stringent constraint on the molecular gas reservoir of a quiescent galaxy beyond $z>1$ and one of the deepest outside the local universe to date. 
This limit of \fgas $\lesssim7\%$, indicates that the gas depletion in this galaxy was effective and nearly complete. As discussed in \S \ref{ancillary}, the few constraints that exist among quenched galaxies beyond $z>1$ collectively indicate a surprising diversity of molecular gas contents, ranging from $f_{H_2}$ between $\lesssim10\%$ \citep{sargent:15} to $\sim16$\% \citep{gobat:18} to as high as $\sim$40\% in two cluster galaxies \citep{hayashi:18, rudnick:17}.  Similar diversity has been observed among quiescent and post-starburst galaxies at low and intermediate redshifts \citep[4-30\%;][]{french:15, suess:17, spilker:18}. This object, combined with the two gas-rich post-starburst galaxies \citet{suess:17}, provides evidence for a similar diversity of gas contents in galaxies at $z>0$ immediately following quenching. The scatter or distribution of molecular gas content with respect to stellar age may encode critical hints towards the physics driving the quenching process in massive galaxies.

We also find that the molecular gas fraction is strongly correlated with the stellar mass surface density at all epochs{, both in the local Universe \citep[see also e.g.,][]{saintonge:11} and since $z\sim2$. \alphaco remains a source of systematic uncertainty, but we note that, for example, adopting lower value of \alphaco for the galaxies with the highest stellar densities \citep[e.g.,][]{bolatto:13} would only strengthen the break between populations in Figure \ref{fig:sigmastar}b. Although idealized merger simulations predict a broad range of \alphaco \citep[e.g.,][]{narayanan:11} as could be relevant for compact merger remnants, it would be very difficult to produce higher \alphaco values unless the galaxies have significantly sub-solar metallicity. We expect the dense galaxies in these samples to be metal-rich given their high stellar masses and therefore posit that adopting a Milky Way \alphaco is a conservative assumption.} This result may be another manifestation 
of the reasonable correlation between stellar and gas structures \citep[e.g.][]{tacconi:13} and the Kennicutt-Schmidt relation \citep{schmidt:59,kennicutt:98}. Perhaps this implies that high stellar densities facilitate efficient star formation and gas consumption. However we note that such correlations may also result from the effects of progenitor biases \citep{lilly:16}. Alternatively, more compact galaxies have higher stellar velocity dispersions and therefore likely host larger supermassive black holes, which may in turn drive stronger feedback \citep[e.g.][]{magorrian:98}. There is a well-established correlation between galaxy density - either on average or within a central region - and stellar populations and ongoing star formation \citep[e.g.,][and references therein]{kauffmannstruct:03,franx:08,cheung:12, fang:13, whitaker:17}. The correlations in Figure \ref{fig:sigmastar}b suggest that this may be related to the gas reservoirs fueling that star formation, but given the sparse sampling of very dense galaxies at high redshift we are limited to speculation. {It is also possible that causality points in the other direction and that the sharp transition in gas fraction is driven by differences in specific star formation rates, which in turn correlate with structures.} We stress that obtaining larger samples will be crucial in quantifying that diversity and exploring correlations with stellar populations and structures, particularly pushing to high redshift where observations probe closer to the quenching epoch for massive galaxies.

\acknowledgements{
This paper makes use of the following ALMA data: ADS/JAO.ALMA\#2015.1.00853.S. ALMA is a partnership of ESO (representing its member states), NSF (USA) and NINS (Japan), together with NRC (Canada), NSC and ASIAA (Taiwan), and KASI (Republic of Korea), in cooperation with the Republic of Chile. The Joint ALMA Observatory is operated by ESO, AUI/NRAO and NAOJ.
The National Radio Astronomy Observatory is a facility of the National Science Foundation operated under cooperative agreement by Associated Universities, Inc. CCW acknowledges support from the National Science Foundation Astronomy and Astrophysics Fellowship grant AST-1701546.
}

\end{document}